\documentstyle[12pt]{article}

\title{A COVARIANT METHOD FOR CALCULATING AMPLITUDES OF PROCESSES
       INVOLVING POLARIZED SPIN $1/2$ PARTICLES. CALCULATION OF
       THE INTERFERENCE TERMS  IN THEIR CROSS SECTIONS}

\author {Alexander~L.~Bondarev
\and \it National Scientific and Educational Center of Particle and
\and \it High Energy Physics attached to Belarusian State University
\and \it M.Bogdanovich str.,153, Minsk 220040, Republic of Belarus
\and \rm e-mail: bondarev@hep.by}

\begin{document}
\maketitle

Published in \\
{\it Teoreticheskaya i Matematicheskaya Fizika},
Vol.96, No.1, P.96 -- 108 (1993) (in Russian)

Translated in \\
{\it Theoretical and Mathematical Physics},
Vol.96, No.1, P.837 -- 844 (1993)  \\

\begin{abstract}
    A covariant method is proposed for calculating the  amplitudes
of  processes  involving  polarized  spin  $1/2$ particles.  It is
suitable  for  calculating  the  interference  terms  in the cross
sections of such processes.   As an illustration,  expressions are
given for  the amplitudes  of electron-electron  scattering in the
lowest  order  of  perturbation  theory  and  expressions  for the
electron current  in the  case of  emission of  two bremsstrahlung
photons in the ultrarelativistic limit.
\end{abstract}

%%%%%%%%%%%%%%%%%%%%%%%%%%%%%%%%%%%%%%%%%%%%%%%%%%%%%%%%%%%%%%%%%%%%%%%%%
%%%%%%%%%%%%%%%%%%%%%%%%%%%%%%%%%%%%%%%%%%%%%%%%%%%%%%%%%%%%%%%%%%%%%%%%%

\section {Introduction}

    In calculations of cross sections involving diagrams of higher
orders (especially when allowance is made for the polarizations of
the participating particles), the need to calculate the traces  of
products of  a large  number of  Dirac $\gamma$  matrices presents
considerable  difficulties.    One  of  the  ways of avoiding such
difficulties  is  to  calculate  directly  the  amplitudes  of the
processes.  In particular, expressions are given in \cite{r1} that
were  obtained  by  the  multiplication  of  $\gamma$ matrices and
bispinors expressed component by  component in definite frames  of
reference.  Because  of computational difficulties,  other authors
too in  later studies  were forced to use such a  device (see, for
example, \cite{r2}, \cite{r3}).

    The  obvious  shortcomings  of  such  an  approach include the
complexity of the calculations and the cumbersome and noncovariant
nature of the results.

    Various authors have attempted the covariant calculation of
amplitudes (see~\cite{r4}~--~\cite{r10}), but the expressions
obtained in their studies could not be used to calculate the
interference terms in the cross sections. We consider below a
general scheme that embraces the results of the cited studies, and
we propose a particularization of the scheme that permits the
calculation of not only amplitudes but also interference terms in
the cross sections.

%%%%%%%%%%%%%%%%%%%%%%%%%%%%%%%%%%%%%%%%%%%%%%%%%%%%%%%%%%%%%%%%%%%%%%%%%
%%%%%%%%%%%%%%%%%%%%%%%%%%%%%%%%%%%%%%%%%%%%%%%%%%%%%%%%%%%%%%%%%%%%%%%%%

\section {General method of covariant calculation}

    In any reaction involving spin $1/2$ particles in the  initial
and final  states, there  is an  even number  $(2N)$ of  fermions.
Therefore, each diagram  contains $N$ fermion  lines that are  not
closed.  In the amplitude of the process, two bispinors correspond
to the  ends of  each line.   For  definiteness, we  shall in what
follows assume  that both  fermions are  particles.   However, the
results  also  hold  when  the  fermions  are antiparticles or one
fermion is a particle and the other an antiparticle.

    In the amplitude of the process, there corresponds to each
line an expression of the form
\begin {equation}
\displaystyle
 M_{12} = \bar{u}_2 Q u_1
\label{e1}
\end {equation}
where
$
\displaystyle
\,
u_1 = u(p_1,n_1)
\,
$
and
$
\displaystyle
\,
u_2 = u(p_2,n_2)
\,
$
are the bispinors for the free particles,
$p_1$ and $p_2$ are the $4$-momenta of the particles,
$n_1$ and $n_2$ are the $4$-vectors that specify the axes of the
spin projections of the particles,
$
\displaystyle
\bar{u}_2 = u^{+}_2 {\gamma}_4
$
and
$
\displaystyle
u^{+} = \tilde{u}^{*}
$
(the asterisk denotes complex conjugation, the tilde matrix
transposition), and $Q$ is the matrix operator that characterizes
the interactions.

    The operator $Q$ can be  expressed as a linear combination  of
products of Dirac $\gamma$ matrices (or contractions of them  with
$4$-vectors)  and may  have an  arbitrary number  of free  Lorentz
indices.

 To calculate $M_{12}$, we use the scheme
\begin {equation}
\begin {array}{l} \displaystyle
M_{12} = \bar{u}_2 Q u_1 = ( \bar{u}_2 Q u_1 )
       { \bar{u}_1 Z u_2 \over \bar{u}_1 Z u_2 }
= { \bar{u}_2 Q u_1 \bar{u}_1 Z u_2 \over \bar{u}_1 Z u_2 }
                        \\[0.5cm] \displaystyle
= { ( Q u_1 \bar{u}_1 Z u_2 \bar{u}_2 )_t \over
                    \bar{u}_1 Z u_2 }
\simeq { ( Q u_1 \bar{u}_1 Z u_2 \bar{u}_2 )_t \over
         | \bar{u}_1 Z u_2 | }
     = { ( Q u_1 \bar{u}_1 Z u_2 \bar{u}_2 )_t \over
 [ ( \bar{Z} u_1 \bar{u}_1 Z u_2 \bar{u}_2 )_t ]^{1/2} }
   = {\cal M}_{12}
\end {array}
\label{e2}
\end {equation}
where $Z$ is an arbitrary $4 \times 4$ matrix,
$$
\displaystyle
\bar{Z} = {\gamma}_4 Z^{+} {\gamma}_4 \; , \;\;
Z^{+} = \tilde{Z}^{*}
$$
($t$ is the symbol of the matrix trace, and the symbol $\simeq$
means "up to a phase factor").

In place of $u\bar{u}$ in (\ref{e2}) projection operators are
substituted. For a massive particle
\begin {equation}
\displaystyle
u(p,n) \bar{u}(p,n) = { 1 \over 4m }( m - {\it i} \hat{p} )
             ( 1 + {\it i} \gamma_5 \hat{n} ) = {\cal P} \, ,
\label{e3}
\end {equation}
where
$
\displaystyle
\hat{p} = p_{\mu} \gamma_{\mu}
, \;
p^2 = - m^2
, \;
n^2 = 1
, \;
pn = 0
, \;
\bar{u} u = 1
, \;
\gamma_5 =  \gamma_1 \gamma_2 \gamma_3 \gamma_4
\,
$ and
$m$ is the particle mass [we use a metric in which
$
\displaystyle
a_{\mu} = ( \vec{a} \, , a_4 = {\it i} a_0 ) ,
\;
ab = a_{\mu} b_{\mu} = \vec{a} \vec{b} - a_0 b_0
$
].

 For a massless particle, the projection operator has the form
\begin {equation}
\displaystyle
u_{\pm}(q) \bar{u}_{\pm}(q)
= { 1 \over 4 {\it i} q_0 }
( 1 \mp \gamma_5 ) \hat{q} = {\cal P}_{\pm}
\label{e4}
\end {equation}
where
$
\displaystyle
q^2 = 0
\,
$
and
$
\displaystyle
\bar{u}_{\pm} \gamma_4 u_{\pm} = 1
\,
$
(the signs $\pm$ of ${\cal P}$ correspond to the particle
helicities).

In \cite{r4}, \cite{r5}, a choice proposed for $Z$ was
$
\displaystyle
Z = 1
\, .
$
Another proposal in \cite{r5} was
$
\displaystyle
Z = \gamma_5
\,
$;
in addition, the expressions obtained in  \cite{r6} were given.
The results of \cite{r7}~--~\cite{r9} reduce to
$
\displaystyle
Z = 1 + \gamma_4
\,
$
and the results of \cite{r10} to
$
\displaystyle
Z = m - {\it i} \hat{r}
$
    ($r$ is an arbitrary $4$-momentum such that $ r^2 = -m^2 $);
the $4$-vectors used in \cite{r10} to specify the axes of the spin
projections were
$$
\displaystyle
n_1 = { m^2 p_2 + ( p_1 p_2 ) p_1 \over
   m [ ( p_1 p_2 )^2 - m^4 ]^{1/2} }
\;\;\; , \;\;\;\;
n_2 = - { m^2 p_1 + ( p_1 p_2 ) p_2 \over
   m [ ( p_1 p_2 )^2 - m^4 ]^{1/2} }
\;\;.
$$

    However, as can be seen from (\ref{e2}), all the expressions
obtained for $M_{12}$ are known up to a phase factor that depends
on
$
\displaystyle
{\bar u}_1 Z u_2
\, .
$

   We note in passing that
\begin {equation}
\begin {array}{l} \displaystyle
( M_{12} )^{*}
\simeq { [ ( Q u_1 {\bar u}_1 Z u_2 {\bar u}_2 )_t ]^{*} \over
 [ ( \bar{Z} u_1 {\bar u}_1 Z u_2 {\bar u}_2 )_t ]^{1/2} }
 = { ( \bar{u}_2 Q u_1 \bar{u}_1 Z u_2 )^{*} \over
 [ ( \bar{Z} u_1 {\bar u}_1 Z u_2 {\bar u}_2 )_t ]^{1/2} }
                        \\[0.5cm] \displaystyle
 = {  \bar{u}_2 \bar{Z} u_1 \bar{u}_1 \bar{Q} u_2  \over
 [ ( \bar{Z} u_1 {\bar u}_1 Z u_2 {\bar u}_2 )_t ]^{1/2} }
 = { ( \bar{Z} u_1 {\bar u}_1 \bar{Q}  u_2 {\bar u}_2 )_t  \over
 [ ( \bar{Z} u_1 {\bar u}_1 Z u_2 {\bar u}_2 )_t ]^{1/2} }
=  ( {\cal M}_{12} )^{*}
\;\; ,
\end {array}
\label{e5}
\end {equation}

  The presence of the unknown phase factors makes it impossible in
the general case to use (\ref{e2}) and (\ref{e5}) to calculate
amplitudes of processes that take place through several channels,
since in this case errors in calculating the interference terms in
the cross sections of such processes are possible.

%%%%%%%%%%%%%%%%%%%%%%%%%%%%%%%%%%%%%%%%%%%%%%%%%%%%%%%%%%%%%%%%%%%%%%%%%
%%%%%%%%%%%%%%%%%%%%%%%%%%%%%%%%%%%%%%%%%%%%%%%%%%%%%%%%%%%%%%%%%%%%%%%%%

\section {Calculation of interference terms in the cross sections}

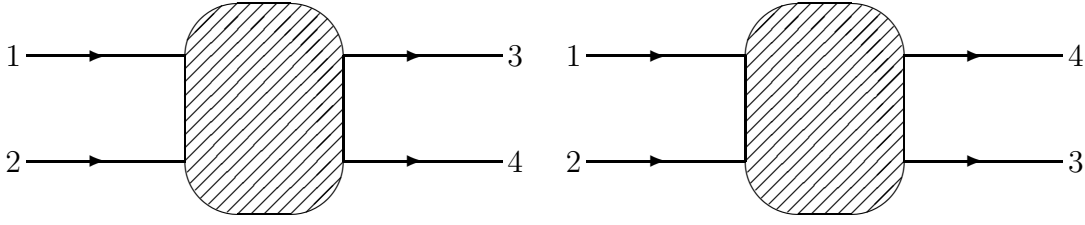
\begin{figure}[ht]
%\vspace{-1cm}
%\hspace{-1cm}
\begin{tabular}{cc}
\begin{picture}(200,100)
\put(100,50){\oval(60,80)}
\put(71,76){\line(1,1){13.}}
\put(70,70){\line(1,1){20.}}
\put(70,65){\line(1,1){25.}}
\put(70,60){\line(1,1){30.}}
\put(70,55){\line(1,1){35.}}
\put(70,50){\line(1,1){40.}}
\put(70,45){\line(1,1){44.}}
\put(70,40){\line(1,1){48.}}
\put(70,35){\line(1,1){51.}}
\put(70,30){\line(1,1){54.}}
\put(71,26){\line(1,1){55.}}
\put(72,22){\line(1,1){56.}}
\put(74,19){\line(1,1){55.}}
\put(76,16){\line(1,1){54.3}}
\put(79,14){\line(1,1){51.}}
\put(82,12){\line(1,1){48.}}
\put(86,11){\line(1,1){44.}}
\put(90,10){\line(1,1){40.}}
\put(95,10){\line(1,1){35.}}
\put(100,10){\line(1,1){30.}}
\put(105,10){\line(1,1){25.}}
\put(110,10){\line(1,1){20.}}
\put(116,11){\line(1,1){13.}}
\put(05,30){\makebox(0,0){$ 2 $}}
\put(05,70){\makebox(0,0){$ 1 $}}
\put(195,30){\makebox(0,0){$ 4 $}}
\put(195,70){\makebox(0,0){$ 3 $}}
\thicklines
\put(10,30){\line(1,0){60.}}
\thicklines
\put(10,70){\line(1,0){60.}}
\thicklines
\put(130,30){\line(1,0){60.}}
\thicklines
\put(130,70){\line(1,0){60.}}
\put(40,30){\vector(1,0){0.}}
\put(40,70){\vector(1,0){0.}}
\put(160,30){\vector(1,0){0.}}
\put(160,70){\vector(1,0){0.}}
\end{picture}
&
\begin{picture}(200,100)
\put(100,50){\oval(60,80)}
\put(71,76){\line(1,1){13.}}
\put(70,70){\line(1,1){20.}}
\put(70,65){\line(1,1){25.}}
\put(70,60){\line(1,1){30.}}
\put(70,55){\line(1,1){35.}}
\put(70,50){\line(1,1){40.}}
\put(70,45){\line(1,1){44.}}
\put(70,40){\line(1,1){48.}}
\put(70,35){\line(1,1){51.}}
\put(70,30){\line(1,1){54.}}
\put(71,26){\line(1,1){55.}}
\put(72,22){\line(1,1){56.}}
\put(74,19){\line(1,1){55.}}
\put(76,16){\line(1,1){54.3}}
\put(79,14){\line(1,1){51.}}
\put(82,12){\line(1,1){48.}}
\put(86,11){\line(1,1){44.}}
\put(90,10){\line(1,1){40.}}
\put(95,10){\line(1,1){35.}}
\put(100,10){\line(1,1){30.}}
\put(105,10){\line(1,1){25.}}
\put(110,10){\line(1,1){20.}}
\put(116,11){\line(1,1){13.}}
\put(05,30){\makebox(0,0){$ 2 $}}
\put(05,70){\makebox(0,0){$ 1 $}}
\put(195,30){\makebox(0,0){$ 3 $}}
\put(195,70){\makebox(0,0){$ 4 $}}
\thicklines
\put(10,30){\line(1,0){60.}}
\thicklines
\put(10,70){\line(1,0){60.}}
\thicklines
\put(130,30){\line(1,0){60.}}
\thicklines
\put(130,70){\line(1,0){60.}}
\put(40,30){\vector(1,0){0.}}
\put(40,70){\vector(1,0){0.}}
\put(160,30){\vector(1,0){0.}}
\put(160,70){\vector(1,0){0.}}
\end{picture}
\end{tabular}
\caption{General form the diagrams for a process that
proceeds through two different channels}
\label{Fg1}
\end{figure}

    We consider in general form the diagrams for a process that
proceeds through two different channels (see Fig.\ref{Fg1}).
%%%%%%%%%%%%%%%%%%%%%%%%%%%%%%%%%%%%%%%%%%%%%%%%%%%%%%%%%%%%%%%%%%%%%%%%%
%%%%%%%%%%%%%%%%%%%%%%%%%%%%%%%%%%%%%%%%%%%%%%%%%%%%%%%%%%%%%%%%%%%%%%%%%

    To the first diagram there corresponds the expression
$$
\displaystyle
M = ( \bar{u}_3 A u_1 ) ( \bar{u}_4 B u_2 ) = M_{13} M_{24}
\, ,
$$
and to the second
$$
\displaystyle
 M'= ( \bar{u}_4 C u_1 ) ( \bar{u}_3 D u_2 ) = M_{14} M_{23}
\, ,
$$
  where $A$, $B$, $C$, $D$ are arbitrary matrix operators,
$$
\displaystyle
( M')^{*} = ( \bar{u}_1 \bar{C} u_4 ) ( \bar{u}_2 \bar{D} u_3 )
\, .
$$

    Difficulties arise in the calculation of interference
expressions of the form
\begin {equation}
\begin {array}{c} \displaystyle
M ( M')^{*}
 = ( \bar{u}_3 A u_1 ) ( \bar{u}_4 B u_2 )
   ( \bar{u}_1 \bar{C} u_4 )
   ( \bar{u}_2 \bar{D} u_3 )
                        \\[0.5cm] \displaystyle
= [ A u_1 \bar{u}_1 \bar{C} u_4 \bar{u}_4
    B u_2 \bar{u}_2 \bar{D} u_3 \bar{u}_3 ]_t
\; .
\end {array}
\label{e6}
\end {equation}
To consider (\ref{e6}), we need the following identity:
\begin {equation}
\displaystyle
[ A_1 u_1 \bar{u}_1 A_2 u_3 \bar{u}_3 ]_t
[ Z_1 u_1 \bar{u}_1 Z_2 u_3 \bar{u}_3 ]_t
\equiv [ A_1 u_1 \bar{u}_1 Z_2 u_3 \bar{u}_3 ]_t
       [ Z_1 u_1 \bar{u}_1 A_2 u_3 \bar{u}_3 ]_t
\, ,
\label{e7}
\end {equation}
where $A_1$, $A_2$, $Z_1$, $Z_2$ are arbitrary $ 4 \times 4 $
matrices.

    The validity of the identity (\ref{e7}) becomes obvious if
each of its sides is rewritten as a product of four currents.

    We apply the consequence of (\ref{e7}):
$$
\displaystyle
[ A_1 u_1 \bar{u}_1 A_2 u_3 \bar{u}_3 ]_t
= { [ A_1 u_1 \bar{u}_1 Z_2 u_3 \bar{u}_3 ]_t
    [ Z_1 u_1 \bar{u}_1 A_2 u_3 \bar{u}_3 ]_t  \over
    [ Z_1 u_1 \bar{u}_1 Z_2 u_3 \bar{u}_3 ]_t }
\, .
$$
    Let
$
\displaystyle
A_1 = A
\, , \;
A_2 = [ \bar{C} u_4 \bar{u}_4 B u_2 \bar{u}_2 \bar{D} ]
\, , \;
Z_2 = Z
\, , \;
Z_2 = \bar{Z}
\, .
$
   Then
\begin {equation}
\begin {array}{r} \displaystyle
M(M')^{*}
= { [ A u_1 \bar{u}_1 Z u_3 \bar{u}_3 ]_t
    [ \bar{Z} u_1 \bar{u}_1 \bar{C} u_4 \bar{u}_4
      B u_2 \bar{u}_2 \bar{D} u_3 \bar{u}_3 ]_t  \over
    [ \bar{Z} u_1 \bar{u}_1 Z u_3 \bar{u}_3 ]_t }
                        \\[0.5cm] \displaystyle
= { [ A u_1 \bar{u}_1 Z u_3 \bar{u}_3 ]_t \over
    [ \bar{Z} u_1 \bar{u}_1 Z u_3 \bar{u}_3 ]_t }
    [ \bar{C} u_4 \bar{u}_4 B u_2 \bar{u}_2
      \bar{D} u_3 \bar{u}_3 \bar{Z} u_1 \bar{u}_1]_t
\, .
\end {array}
\label{e8}
\end {equation}

   Similarly, for the second factor in (\ref{e8})
\begin {equation}
\begin {array}{r} \displaystyle
    [ \bar{C} u_4 \bar{u}_4 B u_2 \bar{u}_2
      \bar{D} u_3 \bar{u}_3 \bar{Z} u_1 \bar{u}_1]_t
= { [ \bar{C} u_4 \bar{u}_4 \bar{X} u_1 \bar{u}_1 ]_t
    [ X u_4 \bar{u}_4 B u_2 \bar{u}_2
      \bar{D} u_3 \bar{u}_3 \bar{Z} u_1 \bar{u}_1 ]_t  \over
    [ X u_4 \bar{u}_4 \bar{X} u_1 \bar{u}_1 ]_t }
                        \\[0.5cm] \displaystyle
= { [ \bar{X} u_1 \bar{u}_1 \bar{C} u_4 \bar{u}_4 ]_t \over
    [ \bar{X} u_1 \bar{u}_1 X u_4 \bar{u}_4 ]_t }
    [ B u_2 \bar{u}_2 \bar{D} u_3 \bar{u}_3
      \bar{Z} u_1 \bar{u}_1 X u_4 \bar{u}_4]_t
\, ,
\end {array}
\label{e9}
\end {equation}
   and also
\begin {equation}
\begin {array}{r} \displaystyle
    [ B u_2 \bar{u}_2 \bar{D} u_3 \bar{u}_3
      \bar{Z} u_1 \bar{u}_1 X u_4 \bar{u}_4]_t
= { [ B u_2 \bar{u}_2 Y u_4 \bar{u}_4 ]_t
    [ \bar{Y} u_2 \bar{u}_2 \bar{D} u_3 \bar{u}_3
      \bar{Z} u_1 \bar{u}_1 X u_4 \bar{u}_4 ]_t  \over
    [ \bar{Y} u_2 \bar{u}_2 Y u_4 \bar{u}_4 ]_t }
                        \\[0.5cm] \displaystyle
= { [ B u_2 \bar{u}_2 Y u_4 \bar{u}_4 ]_t \over
    [ \bar{Y} u_2 \bar{u}_2 Y u_4 \bar{u}_4 ]_t }
    [ \bar{D} u_3 \bar{u}_3 \bar{Z} u_1 \bar{u}_1
      X u_4 \bar{u}_4 \bar{Y} u_2 \bar{u}_2]_t
\, .
\end {array}
\label{e10}
\end {equation}
   Finally
\begin {equation}
\begin {array}{r} \displaystyle
    [ \bar{D} u_3 \bar{u}_3 \bar{Z} u_1 \bar{u}_1
      X u_4 \bar{u}_4 \bar{Y} u_2 \bar{u}_2]_t
= { [ \bar{D} u_3 \bar{u}_3 \bar{V} u_2 \bar{u}_2 ]_t
    [ V u_3 \bar{u}_3 \bar{Z} u_1 \bar{u}_1
      X u_4 \bar{u}_4 \bar{Y} u_2 \bar{u}_2 ]_t  \over
    [ V u_3 \bar{u}_3 \bar{V} u_2 \bar{u}_2 ]_t }
                        \\[0.5cm] \displaystyle
= { [ \bar{V} u_2 \bar{u}_2 \bar{D} u_3 \bar{u}_3 ]_t \over
    [ \bar{V} u_2 \bar{u}_2 V u_3 \bar{u}_3 ]_t }
    [ V u_3 \bar{u}_3 \bar{Z} u_1 \bar{u}_1
      X u_4 \bar{u}_4 \bar{Y} u_2 \bar{u}_2]_t
\, .
\end {array}
\label{e11}
\end {equation}
   In obtaining (\ref{e8}) -- (\ref{e11}), we have used when
necessary cyclic permutations of matrices under the trace symbol;
$ \, X$, $Y$, $Z$, $V \, $
are as yet arbitrary $4 \times 4$ matrices.

   Combining (\ref{e6}), (\ref{e8}) -- (\ref{e11}), we obtain
\begin {equation}
\begin {array}{c} \displaystyle
M ( M')^{*}
= { [ A u_1 \bar{u}_1 Z u_3 \bar{u}_3 ]_t \over
    [ \bar{Z} u_1 \bar{u}_1 Z u_3 \bar{u}_3 ]_t }
  { [ \bar{X} u_1 \bar{u}_1 \bar{C} u_4 \bar{u}_4 ]_t \over
    [ \bar{X} u_1 \bar{u}_1 X u_4 \bar{u}_4 ]_t }
             \\           \\     \displaystyle
\times { [ B u_2 \bar{u}_2 Y u_4 \bar{u}_4 ]_t \over
         [ \bar{Y} u_2 \bar{u}_2 Y u_4 \bar{u}_4 ]_t }
  { [ \bar{V} u_2 \bar{u}_2 \bar{D} u_3 \bar{u}_3 ]_t \over
    [ \bar{V} u_2 \bar{u}_2 V u_3 \bar{u}_3 ]_t }
 [ V u_3 \bar{u}_3 \bar{Z} u_1 \bar{u}_1
   X u_4 \bar{u}_4 \bar{Y} u_2 \bar{u}_2 ]_t
              \\          \\    \displaystyle
= { [ A u_1 \bar{u}_1 Z u_3 \bar{u}_3 ]_t \over
  ( [ \bar{Z} u_1 \bar{u}_1 Z u_3 \bar{u}_3 ]_t )^{1/2} }
  { [ \bar{X} u_1 \bar{u}_1 \bar{C} u_4 \bar{u}_4 ]_t \over
  ( [ \bar{X} u_1 \bar{u}_1 X u_4 \bar{u}_4 ]_t )^{1/2} }
  { [ B u_2 \bar{u}_2 Y u_4 \bar{u}_4 ]_t \over
  ( [ \bar{Y} u_2 \bar{u}_2 Y u_4 \bar{u}_4 ]_t )^{1/2} }
  { [ \bar{V} u_2 \bar{u}_2 \bar{D} u_3 \bar{u}_3 ]_t \over
  ( [ \bar{V} u_2 \bar{u}_2 V u_3 \bar{u}_3 ]_t )^{1/2} }
              \\          \\        \displaystyle
\times
  {  [ \bar{Z} u_1 \bar{u}_1 X u_4 \bar{u}_4
       \bar{Y} u_2 \bar{u}_2 V u_3 \bar{u}_3 ]_t \over
   [ ( \bar{Z} u_1 \bar{u}_1 Z u_3 \bar{u}_3 )_t
     ( \bar{X} u_1 \bar{u}_1 X u_4 \bar{u}_4 )_t
     ( \bar{Y} u_2 \bar{u}_2 Y u_4 \bar{u}_4 )_t
    ( \bar{V} u_2 \bar{u}_2 V u_3 \bar{u}_3 )_t ]^{1/2} }
              \\          \\        \displaystyle
= {\cal M}_{13} {\cal M}_{24} ( {\cal M}_{14} )^{*}
                   ( {\cal M}_{23} )^{*} K
\, ,
\end {array}
\label{e12}
\end {equation}
where
$ \, {\cal M}_{13}$, ${\cal M}_{24}$,
$( {\cal M}_{14} )^{*}$, $( {\cal M}_{23} )^{*}$
are given by expressions analogous to (\ref{e2}) and (\ref{e5}),
and the coefficient $K$ is given by
\begin {equation}
\displaystyle
K= { [ \bar{Z} u_1 \bar{u}_1 X u_4 \bar{u}_4
       \bar{Y} u_2 \bar{u}_2 V u_3 \bar{u}_3 ]_t \over
   [ ( \bar{Z} u_1 \bar{u}_1 Z u_3 \bar{u}_3 )_t
     ( \bar{X} u_1 \bar{u}_1 X u_4 \bar{u}_4 )_t
     ( \bar{Y} u_2 \bar{u}_2 Y u_4 \bar{u}_4 )_t
    ( \bar{V} u_2 \bar{u}_2 V u_3 \bar{u}_3 )_t ]^{1/2} }
\, .
\label{e13}
\end {equation}

    Obviously, for correct calculation of the interference
contributions it is necessary to require
$
\displaystyle
   K \equiv 1
\, .
$
    This requirement is satisfied if we choose   \\
$
\displaystyle
Z = X = Y = V = {\cal P} \;\;\;
$
[see (\ref{e3})] $\;\;$
\hspace{2mm} or \hspace{5mm}
$
\displaystyle
Z = X = Y = V = {\cal P}_{\pm} \;\;\;
$
[see (\ref{e4})] ,  \\
since for the projection operators we have the identities
\begin {equation}
\displaystyle
{\cal P} A {\cal P} = [ {\cal P} A ]_t {\cal P}
\;\; ,   \;\;\;\;\;
\bar{\cal P} = {\cal P}
\; ,
\label{e14}
\end {equation}
\begin {equation}
\displaystyle
{\cal P}_{\pm} A {\cal P}_{\pm}
= [ {\cal P}_{\pm} A ]_t {\cal P}_{\pm}
\;\; , \;\;\;\;\;
\bar{\cal P}_{\pm} = {\cal P}_{\pm}
\; .
\label{e15}
\end {equation}

    As an example, we give expressions for amplitudes of processes
involving massless Dirac particles. In this case, the expression
(\ref{e2}) becomes
\begin {equation}
\displaystyle
\bar{u}_{\pm}(p_2) Q u_{\pm}(p_1)
= { {\it i} [ Q \hat{p}_1 \hat{q} \hat{p}_2 ( 1 \pm \gamma_5 ) ]_t
    \over 8 [ (p_1)_0 (p_2)_0  (q p_1) (q  p_2) ]^{1/2}  }
\; ,
\label{e16}
\end {equation}
where
$
\displaystyle
Z = { 1 \over 4 {\it i} q_0 } ( 1 \pm \gamma_5 ) \hat{q}
=  {\cal P}_{\mp}
\;\;  ,  \;\;\;
q^2  =  0.
$

    The massless $4$-vector $q$ can be arbitrary, but it must be
the same for all the considered fermion lines that are not closed
in the diagrams. Note that in (\ref{e16}) and similar expressions
we shall in what follows use the equals sign instead of the symbol
$ \simeq $ since no problems can now arise with the phase factors.

   Further,
\begin {equation}
\displaystyle
\bar{u}_{\pm}(p_2) Q u_{\mp}(p_1)
= - {  [ Q \hat{p}_1
( m + \gamma_5 \hat{n} \hat{p} ) \hat{p}_2
( 1 \pm \gamma_5 ) ]_t \over
8 \Bigl\{ (p_1)_0 (p_2)_0 [ ( p p_1 ) \pm m ( n p_1 ) ]
         [ (p p_2) \mp m (n p_2) ] \Bigr\}^{1/2} }
\; ,
\label{e17}
\end {equation}
where
$
\displaystyle
Z = { 1 \over 4m } ( m - {\it i} \hat{p} )
( 1 + {\it i} \gamma_5 \hat{n} ) = {\cal P} ,
\;\;
p^2 = - m^2 , \;\; n^2 = 1 ,  \;\; pn = 0 .
$
    With regard to the $4$-vectors $p$ and $n$, the same remark
holds as for the vector $q$ in (\ref{e16}).

    In the last case, it is not possible to use as $Z$ the simpler
operator ${\cal P}_{\pm}$ because both numerator and denominator
would then be identically equal to zero.

    If for certain values of  $p_1$ and $p_2$ the denominators  in
(\ref{e16}) or (\ref{e17}) vanish in the numerical calculation, it
is sufficient to change the  values of the arbitrary vectors  that
appear   in   these   expressions,   namely,   $q$   or  $p$,  $n$
(simultaneously for all considered lines of the diagrams).

    As  we  have  already   noted,  our  method  can   be  readily
generalized to include antiparticles.  For this, it is  sufficient
in (\ref{e2}) to replace the particle projection operators by  the
antiparticle analogs.  For  example, suppose we  are interested in
$\bar{v}_2 Q u_1$ , where  $v_2$ is a free antiparticle  bispinor.
Then
$$
\displaystyle
\bar{v}_2 Q u_1 = { ( Q u_1 \bar{u}_1 Z v_2 \bar{v}_2 )_t
\over
[ ( \bar{Z} u_1 \bar{u}_1 Z v_2 \bar{v}_2 )_t ]^{1/2} }
$$
where
$$
\displaystyle
v(p,n) \bar{v}(p,n)
= - { 1 \over 4m }
  ( m + {\it i} \hat{p} ) ( 1 + {\it i} \gamma_5 \hat{n} )
$$
for a massive antiparticle or
$$
\displaystyle
v_{\pm}(q) \bar{v}_{\pm}(q)
= { 1 \over 4 {\it i} q_0 } ( 1 \pm \gamma_5 ) \hat{q}
$$
for a massless antiparticle. For $Z$, we still use (\ref{e3}) or
(\ref{e4}).

   In conclusion, we note that the previously proposed direct
methods of calculating the amplitudes of processes using
$
\displaystyle
Z = 1 \, , \;  \gamma_5 \, , \; 1 + \gamma_4 \, , \;
m - {\it i} \hat{r}
$
do not permit correct calculation of the interference terms in the
cross section; for when they are used, the coefficient $K$, which
is determined by the expression (\ref{e13}) and takes into account
the phase factors of all the fermion lines that are not closed in
the diagrams, is not equal to $1$ (this is readily demonstrated by
considering, for example, processes involving massless particles).
                     \\

%%%%%%%%%%%%%%%%%%%%%%%%%%%%%%%%%%%%%%%%%%%%%%%%%%%%%%%%%%%%%%%%%%%%%%%%%
%%%%%%%%%%%%%%%%%%%%%%%%%%%%%%%%%%%%%%%%%%%%%%%%%%%%%%%%%%%%%%%%%%%%%%%%%

  {\bf APPENDIX 1}

\begin{figure}[ht]
%\vspace{-1cm}
%\hspace{-1cm}
\begin{tabular}{cc}
\begin{picture}(200,100)
\put(05,30){\makebox(0,0){$ 2 $}}
\put(05,70){\makebox(0,0){$ 1 $}}
\put(195,30){\makebox(0,0){$ 4 $}}
\put(195,70){\makebox(0,0){$ 3 $}}
%
%\thicklines
\put(10,30){\line(1,0){180.}}
%\thicklines
\put(10,70){\line(1,0){180.}}
\put(100,30){\circle*{2}}
\put(100,70){\circle*{2}}
\put(40,30){\vector(1,0){0.}}
\put(40,70){\vector(1,0){0.}}
\put(160,30){\vector(1,0){0.}}
\put(160,70){\vector(1,0){0.}}
\multiput(100,32)(0,8){5}{\oval(4.0,4.0)[r]}
\multiput(100,36)(0,8){5}{\oval(4.0,4.0)[l]}
\end{picture}
&
\begin{picture}(200,100)
\put(05,30){\makebox(0,0){$ 2 $}}
\put(05,70){\makebox(0,0){$ 1 $}}
\put(195,30){\makebox(0,0){$ 3 $}}
\put(195,70){\makebox(0,0){$ 4 $}}
%
%\thicklines
\put(10,30){\line(1,0){180.}}
%\thicklines
\put(10,70){\line(1,0){180.}}
\put(100,30){\circle*{2}}
\put(100,70){\circle*{2}}
\put(40,30){\vector(1,0){0.}}
\put(40,70){\vector(1,0){0.}}
\put(160,30){\vector(1,0){0.}}
\put(160,70){\vector(1,0){0.}}
\multiput(100,32)(0,8){5}{\oval(4.0,4.0)[r]}
\multiput(100,36)(0,8){5}{\oval(4.0,4.0)[l]}
\end{picture}
\end{tabular}
\caption{The Feynman diagrams for electron-electron scattering
in the lowest order of perturbation theory}
\label{Fg2}
\end{figure}
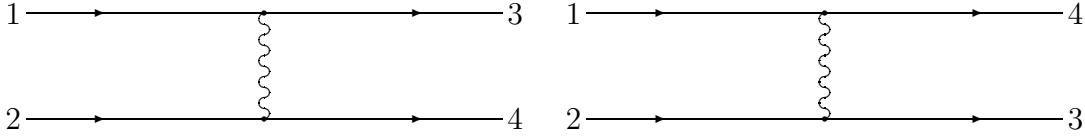

    As an illustration of the application of the method to the
calculation of amplitudes of processes involving massive Dirac
particles, we consider electron-electron scattering in the lowest
order of perturbation theory. To this process there correspond the
two Feynman diagrams shown in Fig.\ref{Fg2}.
%%%%%%%%%%%%%%%%%%%%%%%%%%%%%%%%%%%%%%%%%%%%%%%%%%%%%%%%%%%%%%%%%%%%%%%%%
%%%%%%%%%%%%%%%%%%%%%%%%%%%%%%%%%%%%%%%%%%%%%%%%%%%%%%%%%%%%%%%%%%%%%%%%%

   For calculations of amplitudes of processes involving massive
Dirac particles, formula (\ref{e2}) becomes
\begin {equation}
\displaystyle
M_{12} = \bar{u}_2 Q u_1
     = { ( Q u_1 \bar{u}_1 {\cal P}_{\pm} u_2 \bar{u}_2 )_t
      \over
 [ ( {\cal P}_{\pm} u_1 \bar{u}_1
     {\cal P}_{\pm} u_2 \bar{u}_2 )_t ]^{1/2} }
\, ,
\label{ea1.1}
\end {equation}
and therefore to the electron current of the top line of the first
diagram there will correspond the expression
$$
\begin {array}{l} \displaystyle
(J_{13})_{\mu} = \bar{u}_3 \gamma_{\mu} u_1
 = { [ \gamma_{\mu} {\cal P}_1 {\cal P}_{+} {\cal P}_3 ]_t
      \over
 ( [ {\cal P}_{+} {\cal P}_1
     {\cal P}_{+} {\cal P}_3 ]_t )^{1/2} }
                        \\[0.5cm] \displaystyle
= { \left[ \gamma_{\mu}
  { 1 \over 4m } ( m - {\it i} \hat{p}_1 )
             ( 1 + {\it i} \gamma_5 \hat{n}_1 )
  { 1 \over 4 {\it i} q_0 } ( 1 - \gamma_5 ) \hat{q}
  { 1 \over 4m } ( m - {\it i} \hat{p}_3 )
             ( 1 + {\it i} \gamma_5 \hat{n}_3 ) \right]_t
                    \over
   \left( \left[
  { ( 1 - \gamma_5 ) \hat{q} \over 4 {\it i} q_0 }
  { ( m - {\it i} \hat{p}_1 )
             ( 1 + {\it i} \gamma_5 \hat{n}_1 ) \over 4m }
  { ( 1 - \gamma_5 ) \hat{q} \over 4 {\it i} q_0 }
  { ( m - {\it i} \hat{p}_3 )
             ( 1 + {\it i} \gamma_5 \hat{n}_3 ) \over 4m }
             \right]_t \right)^{1/2} }
           \\       \\  \displaystyle
= \alpha^{-1}  ( a q_{\mu} + b {p_1}_{\mu} + c {p_3}_{\mu}
      + d {n_1}_{\mu} + e {n_3}_{\mu} + f_{\mu} )
\end {array}
\, ,
$$
where
$$
\displaystyle
\alpha = 4 {\it i} m [ - ( q p_1 ) + m ( q n_1 ) ]^{1/2}
                     [ - ( q p_3 ) + m ( q n_3 ) ]^{1/2}
\, ,
$$
$$
\displaystyle
a =  [ m^2 + ( p_1 p_3 ) ] [ 1 + ( n_1 n_3 ) ]
  - m ( p_1 n_3 ) - m ( p_3 n_1 ) - ( p_1 n_3 ) ( p_3 n_1 )
\, ,
$$
$$
\displaystyle
b =  m (q n_1 ) + (q n_3 ) [ m + ( p_3 n_1 ) ]
     - (q p_3 ) [ 1 + (n_1 n_3 ) ] - \varepsilon(q, p_3, n_1, n_3)
\, ,
$$
$$
\displaystyle
c =  m (q n_3 ) + (q n_1 ) [ m + ( p_1 n_3 ) ]
     - (q p_1 ) [ 1 + (n_1 n_3 ) ] - \varepsilon(q, p_1, n_1, n_3)
\, ,
$$
$$
\displaystyle
d = -m (q p_1 ) + (q p_3 ) [ m + ( p_1 n_3 ) ]
     - (q n_3 ) [ m^2 + (p_1 p_3 ) ]
     - \varepsilon(q, p_1, p_3, n_3)
\, ,
$$
$$
\displaystyle
e = -m (q p_3 ) + (q p_1 ) [ m + ( p_3 n_1 ) ]
     - (q n_1 ) [ m^2 + (p_1 p_3 ) ]
     - \varepsilon(q, p_1, p_3, n_1)
\, ,
$$
$$
\begin {array}{c} \displaystyle
f_{\mu} = [ m + (p_3 n_1 ) ] \varepsilon(\mu, q, p_1, n_3)
        - [ m + (p_1 n_3 ) ] \varepsilon(\mu, q, p_3, n_1)
                        \\ [0.25cm]  \displaystyle
        - [ 1 + (n_1 n_3 ) ] \varepsilon(\mu, q, p_1, p_3)
        - [ m^2 + (p_1 p_3 ) ] \varepsilon(\mu, q, n_1, n_3)
                        \\ [0.25cm]  \displaystyle
        - m \varepsilon(\mu, q, p_3, n_3)
        + m \varepsilon(\mu, q, p_1, n_1)
\, ,
\end {array}
$$
in which
$
\displaystyle
  \varepsilon(\mu, a, b, c)
= \varepsilon_{\mu \nu \rho \sigma} a_{\nu} b_{\rho} c_{\sigma}
$
is the contraction of the completely antisymmetric Levi-Civita
tensor with the $4$-vectors $a$, $b$ and $c$,
$
\displaystyle
\;
q^2 = 0
\, .
$

   The expression for
$
\displaystyle
\;
(J_{24})_{\mu} = \bar{u}_4 \gamma_{\mu} u_2
\;
$
is obtained from
$
\displaystyle
\;
(J_{13})_{\mu}
\;
$
by replacing the index $3$ by $4$ and $1$ by $2$;
$
\displaystyle
\;
(J_{14})_{\nu} = \bar{u}_4 \gamma_{\nu} u_1
\;
$
is obtained by
$
\displaystyle
\;
(J_{13})_{\mu}
\;
$
by replacing $3$ by $4$ and $\mu$ by $\nu$, and
$
\displaystyle
\;
(J_{23})_{\nu} = \bar{u}_3 \gamma_{\nu} u_2
\;
$
by replacing $1$ by $2$ and $\mu$ by $\nu$, respectively.

    The  expressions  we  have  given  are  covariant  and  permit
numerical  calculations  of  amplitudes.    The  complex   numbers
obtained  in  the  calculation  serve  for  the calculation of the
process cross section.

    In  this  example,  the  calculation  of  the  amplitude is as
laborious as the calculation  of the square  of the modulus of the
matrix element for one diagram but simpler than the calculation of
the  interference  term.    However,  if  the  number  of $\gamma$
matrices   in   the   operator    $Q$   is   increased   by    $N$
[see~(\ref{ea1.1})],   their   number   in   the   numerator    of
(\ref{ea1.1})  increases  only  by  $N$  (at  the  same  time, the
denominator is unchanged), whereas in the construction
$
\displaystyle
\;
[ Q u_1 \bar{u}_1 \bar{Q} u_2 \bar{u}_2 ]_t
\;
$,
    which arises in calculations of the square of the modulus, the
number of $\gamma$ matrices is increased by  $2N$. Since the trace
of product of $2M$ $\gamma$ matrices contains
$
\displaystyle
\;
1 \cdot 3 \cdot 5 \cdot \ldots \cdot ( 2 M - 1 )
\;
$
    terms, we see that the more complicated the process the
greater the gain from calculating it in the method of direct
calculation of the amplitudes. This is also true for processes
involving massless particles.
              \\

%%%%%%%%%%%%%%%%%%%%%%%%%%%%%%%%%%%%%%%%%%%%%%%%%%%%%%%%%%%%%%%%%%%%%%%%%
%%%%%%%%%%%%%%%%%%%%%%%%%%%%%%%%%%%%%%%%%%%%%%%%%%%%%%%%%%%%%%%%%%%%%%%%%

    {\bf APPENDIX 2}

\begin{figure}[ht]
%\vspace{-1cm}
%\hspace{-1cm}
\begin{tabular}{cccccc}
\begin{picture}(70,105)
\put(05,45){\makebox(0,0){$ q $}}
\put(65,45){\makebox(0,0){$ q' $}}
\put(20,100){\makebox(0,0){$ k_1 $}}
\put(35,100){\makebox(0,0){$ k_2 $}}
\put(35,5){\makebox(0,0){$ 1 $}}
\put(5,50){\line(1,0){60.}}
\put(20,50){\circle*{2}}
\put(35,50){\circle*{2}}
\put(50,50){\circle*{2}}
\put(15,50){\vector(1,0){0.}}
\put(60,50){\vector(1,0){0.}}
\put(20,94){\vector(0,1){0.}}
\put(35,94){\vector(0,1){0.}}
\multiput(20,52)(0,8){5}{\oval(4.0,4.0)[r]}
\multiput(20,56)(0,8){5}{\oval(4.0,4.0)[l]}
\multiput(35,52)(0,8){5}{\oval(4.0,4.0)[r]}
\multiput(35,56)(0,8){5}{\oval(4.0,4.0)[l]}
\multiput(50,12)(0,8){5}{\oval(4.0,4.0)[r]}
\multiput(50,16)(0,8){5}{\oval(4.0,4.0)[l]}
\end{picture}
&
\begin{picture}(70,105)
\put(05,45){\makebox(0,0){$ q $}}
\put(65,45){\makebox(0,0){$ q' $}}
\put(20,100){\makebox(0,0){$ k_2 $}}
\put(35,100){\makebox(0,0){$ k_1 $}}
\put(35,5){\makebox(0,0){$ 2 $}}
\put(5,50){\line(1,0){60.}}
\put(20,50){\circle*{2}}
\put(35,50){\circle*{2}}
\put(50,50){\circle*{2}}
\put(15,50){\vector(1,0){0.}}
\put(60,50){\vector(1,0){0.}}
\put(20,94){\vector(0,1){0.}}
\put(35,94){\vector(0,1){0.}}
\multiput(20,52)(0,8){5}{\oval(4.0,4.0)[r]}
\multiput(20,56)(0,8){5}{\oval(4.0,4.0)[l]}
\multiput(35,52)(0,8){5}{\oval(4.0,4.0)[r]}
\multiput(35,56)(0,8){5}{\oval(4.0,4.0)[l]}
\multiput(50,12)(0,8){5}{\oval(4.0,4.0)[r]}
\multiput(50,16)(0,8){5}{\oval(4.0,4.0)[l]}
\end{picture}
&
\begin{picture}(70,105)
\put(05,45){\makebox(0,0){$ q $}}
\put(65,45){\makebox(0,0){$ q' $}}
\put(20,100){\makebox(0,0){$ k_1 $}}
\put(50,100){\makebox(0,0){$ k_2 $}}
\put(35,5){\makebox(0,0){$ 3 $}}
\put(5,50){\line(1,0){60.}}
\put(20,50){\circle*{2}}
\put(35,50){\circle*{2}}
\put(50,50){\circle*{2}}
\put(15,50){\vector(1,0){0.}}
\put(60,50){\vector(1,0){0.}}
\put(20,94){\vector(0,1){0.}}
\put(50,94){\vector(0,1){0.}}
\multiput(20,52)(0,8){5}{\oval(4.0,4.0)[r]}
\multiput(20,56)(0,8){5}{\oval(4.0,4.0)[l]}
\multiput(50,52)(0,8){5}{\oval(4.0,4.0)[r]}
\multiput(50,56)(0,8){5}{\oval(4.0,4.0)[l]}
\multiput(35,12)(0,8){5}{\oval(4.0,4.0)[r]}
\multiput(35,16)(0,8){5}{\oval(4.0,4.0)[l]}
\end{picture}
&
\begin{picture}(70,105)
\put(05,45){\makebox(0,0){$ q $}}
\put(65,45){\makebox(0,0){$ q' $}}
\put(20,100){\makebox(0,0){$ k_2 $}}
\put(50,100){\makebox(0,0){$ k_1 $}}
\put(35,5){\makebox(0,0){$ 4 $}}
\put(5,50){\line(1,0){60.}}
\put(20,50){\circle*{2}}
\put(35,50){\circle*{2}}
\put(50,50){\circle*{2}}
\put(15,50){\vector(1,0){0.}}
\put(60,50){\vector(1,0){0.}}
\put(20,94){\vector(0,1){0.}}
\put(50,94){\vector(0,1){0.}}
\multiput(20,52)(0,8){5}{\oval(4.0,4.0)[r]}
\multiput(20,56)(0,8){5}{\oval(4.0,4.0)[l]}
\multiput(50,52)(0,8){5}{\oval(4.0,4.0)[r]}
\multiput(50,56)(0,8){5}{\oval(4.0,4.0)[l]}
\multiput(35,12)(0,8){5}{\oval(4.0,4.0)[r]}
\multiput(35,16)(0,8){5}{\oval(4.0,4.0)[l]}
\end{picture}
&
\begin{picture}(70,105)
\put(05,45){\makebox(0,0){$ q $}}
\put(65,45){\makebox(0,0){$ q' $}}
\put(35,100){\makebox(0,0){$ k_1 $}}
\put(50,100){\makebox(0,0){$ k_2 $}}
\put(35,5){\makebox(0,0){$ 5 $}}
\put(5,50){\line(1,0){60.}}
\put(20,50){\circle*{2}}
\put(35,50){\circle*{2}}
\put(50,50){\circle*{2}}
\put(15,50){\vector(1,0){0.}}
\put(60,50){\vector(1,0){0.}}
\put(35,94){\vector(0,1){0.}}
\put(50,94){\vector(0,1){0.}}
\multiput(35,52)(0,8){5}{\oval(4.0,4.0)[r]}
\multiput(35,56)(0,8){5}{\oval(4.0,4.0)[l]}
\multiput(50,52)(0,8){5}{\oval(4.0,4.0)[r]}
\multiput(50,56)(0,8){5}{\oval(4.0,4.0)[l]}
\multiput(20,12)(0,8){5}{\oval(4.0,4.0)[r]}
\multiput(20,16)(0,8){5}{\oval(4.0,4.0)[l]}
\end{picture}
&
\begin{picture}(70,105)
\put(05,45){\makebox(0,0){$ q $}}
\put(65,45){\makebox(0,0){$ q' $}}
\put(35,100){\makebox(0,0){$ k_2 $}}
\put(50,100){\makebox(0,0){$ k_1 $}}
\put(35,5){\makebox(0,0){$ 6 $}}
\put(5,50){\line(1,0){60.}}
\put(20,50){\circle*{2}}
\put(35,50){\circle*{2}}
\put(50,50){\circle*{2}}
\put(15,50){\vector(1,0){0.}}
\put(60,50){\vector(1,0){0.}}
\put(35,94){\vector(0,1){0.}}
\put(50,94){\vector(0,1){0.}}
\multiput(35,52)(0,8){5}{\oval(4.0,4.0)[r]}
\multiput(35,56)(0,8){5}{\oval(4.0,4.0)[l]}
\multiput(50,52)(0,8){5}{\oval(4.0,4.0)[r]}
\multiput(50,56)(0,8){5}{\oval(4.0,4.0)[l]}
\multiput(20,12)(0,8){5}{\oval(4.0,4.0)[r]}
\multiput(20,16)(0,8){5}{\oval(4.0,4.0)[l]}
\end{picture}

\end{tabular}
\caption{The Feynman diagrams for the electron current in the
case of emission of two photons}
\label{Fg3}
\end{figure}
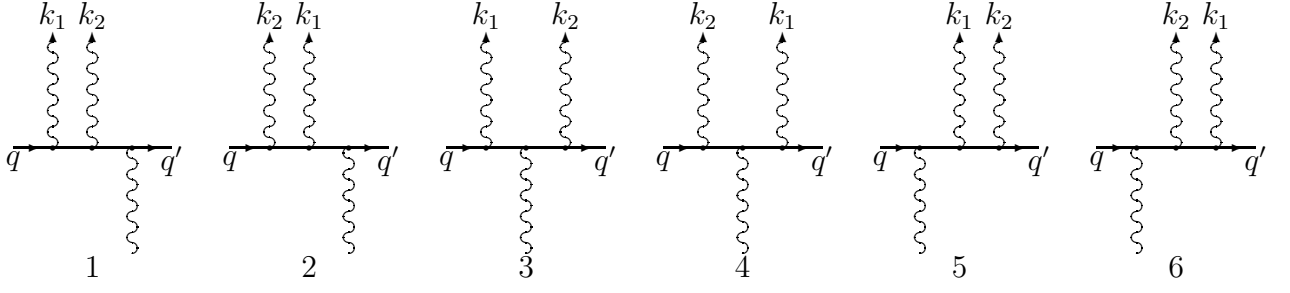

   We consider expressions for the electron current in the case
   of emission of two photons in the ultrarelativistic limit
($
\displaystyle
m_e = 0
\,
$).
The Feynman diagrams are given in Fig.\ref{Fg3}.
%%%%%%%%%%%%%%%%%%%%%%%%%%%%%%%%%%%%%%%%%%%%%%%%%%%%%%%%%%%%%%%%%%%%%%%%%
%%%%%%%%%%%%%%%%%%%%%%%%%%%%%%%%%%%%%%%%%%%%%%%%%%%%%%%%%%%%%%%%%%%%%%%%%

    We take the polarization vectors of a photon with helicity
$\lambda$ in the form (see \cite{r11})
$$
\begin {array}{c} \displaystyle
\hat{e}_{\lambda}(k_i) = 2 N_i
[ \hat{k}_i \hat{q}' \hat{q} \omega_{-\lambda}
- \hat{q}' \hat{q} \hat{k}_i \omega_{\lambda}]
\;\; , \;\;\;\;\;\;\;
\omega_{\pm} = { 1 \over 2 }( 1 \pm \gamma_5 )
\;\; ,
                        \\[0.5cm] \displaystyle
N_i = [ - 16 (q q') (q k_i) (q' k_i) ]^{-1/2}
\; , \;\;\;
i = 1, 2
\;\; .
\end {array}
$$
In the considered approximation, the electrons being massless,
$$
\displaystyle
\omega_{\mp} u_{\pm} = u_{\pm}
\;\; , \;\;
\bar{u}_{\pm} \omega _{\pm} = \bar{u}_{\pm}
\;\; , \;\;
\bar{u}_{\pm}(q') \hat{q}' = 0
\;\; , \;\;
\hat{q} u_{\pm}(q) = 0
\;\; .
$$

   Using the above relations, we obtain
\begin {equation}
\begin {array}{c} \displaystyle
\bar{u}_{\pm}(q') \hat{e}_{\pm}(k_i)
= \bar{u}_{\pm}(q') \omega_{\pm}
2 N_i [ \hat{k}_i \hat{q}' \hat{q} \omega_{\mp}
      - \hat{q}' \hat{q} \hat{k}_i \omega_{\pm} ]
                        \\[0.5cm] \displaystyle
= 2 N_i \bar{u}_{\pm}(q') \hat{k}_i \hat{q}' \hat{q}
= 4 N_i (q' k_i) \bar{u}_{\pm}(q') \hat{q}
\;\; ,
\end {array}
\label{ea2.1}
\end {equation}
\begin {equation}
\begin {array}{c} \displaystyle
\hat{e}_{\mp}(k_i) u_{\pm}(q)
= 2 N_i [ \hat{k}_i \hat{q}' \hat{q} \omega_{\pm}
        - \hat{q}' \hat{q} \hat{k}_i \omega_{\mp} ]
          \omega_{\mp} u_{\pm}(q)
                        \\[0.5cm] \displaystyle
= - 2 N_i \hat{q}' \hat{q} \hat{k}_i u_{\pm}(q)
= - 4 N_i (q k_i) \hat{q}' u_{\pm}(q)
\;\; ,
\end {array}
\label{ea2.2}
\end {equation}

\begin {equation}
\displaystyle
\bar{u}_{\pm}(q') \hat{e}_{\mp}(k_i)
            = \hat{e}_{\pm}(k_i) u_{\pm}(q) = 0
\;\; .
\label{ea2.3}
\end {equation}

    In the expressions given below for the currents
$
\displaystyle
J_{a \mu}( \lambda_1, \lambda_2, \lambda_3, \lambda_4 )
\,
$,
$a$ denotes the diagram number,
$\lambda_1$, $\lambda_2$, $\lambda_3$, $\lambda_4$
are the helicities of the final electron, $\gamma_1$, $\gamma_2$
and of the initial electron, respectively.

Using Eqs. (\ref{e16}) and (\ref{ea2.1}) -- (\ref{ea2.3}), we
obtain
$$
\begin {array}{c} \displaystyle
J_{\mu}(\pm, \pm, \pm, \pm)
= J_{5 \mu}(\pm, \pm, \pm, \pm) + J_{6 \mu}(\pm, \pm, \pm, \pm)
                        \\[0.5cm] \displaystyle
= \bar{u}_{\pm}(q') \hat{e}_{\pm}(k_2)
 { \hat{q}' + \hat{k}_2 \over 2 (q' k_2) } \hat{e}_{\pm}(k_1)
 { \hat{q}' + \hat{k}_1 + \hat{k}_2 \over
            2 ( k_1 k_2 + q' k_1 + q' k_2 ) }
 \gamma_{\mu} u_{\pm}(q) + ( k_1 \leftrightarrow k_2 )
                        \\[0.5cm] \displaystyle
= 8 N_1 N_2 (q q') \bar{u}_{\pm}(q') ( 1 \pm \gamma_5 )
[ ( q q' + q k_1 + q k_2 ) \gamma_{\mu}
- q_{\mu} ( \hat{k}_1 + \hat{k}_2 ) ] u_{\pm}(q)
                        \\[0.5cm] \displaystyle
= 8 N_1 N_2 (q q')
{ {\it i} \over [ q_0 q'_0 (q p) (q' p) ]^{1/2} }
\Bigl\{ [ q'_{\mu} (q p) - p_{\mu} (q q') ]
       ( q q' + q k_1 + q k_2 )
                        \\[0.5cm] \displaystyle
+ q_{\mu} [ (q q') ( q' p + p k_1 + p k_2 )
          - (q p) ( q' k_1 + q' k_2 ) ]
\mp (q p) \varepsilon ( \mu, q, q', k_1 + k_2 )
                        \\[0.5cm] \displaystyle
\mp (q q') \varepsilon ( \mu, q, q' + k_1 + k_2, p ) \Bigr\}
\;\; .
\end {array}
$$
     Similarly
$$
\begin {array}{c} \displaystyle
J_{\mu}( \pm, \mp, \mp, \pm )
= J_{1 \mu}( \pm, \mp, \mp, \pm ) + J_{2 \mu}( \pm, \mp, \mp, \pm )
                        \\[0.5cm] \displaystyle
= 8 N_1 N_2 (q q')
{ {\it i} \over [ q_0 q'_0 (q p) (q' p) ]^{1/2} }
\Bigl\{ [ q_{\mu} (q' p) - p_{\mu} (q q') ]
( q q'- q' k_1 - q' k_2 )
                        \\[0.5cm] \displaystyle
+ {q'}_{\mu} [ ( q q') ( q p - p k_1 - p k_2 )
+ (q'p) ( q k_1 + q k_2 ) ]
\pm (q' p) \varepsilon ( \mu, q, q', k_1 + k_2 )
                        \\[0.5cm] \displaystyle
\mp (q q') \varepsilon (\mu, q - k_1 - k_2, q', p ) \Bigr\}
\;\; ,
\end {array}
$$

$$
\displaystyle
J_{\mu}( \pm, \pm, \mp, \pm ) = J_{2 \mu}( \pm, \pm, \mp, \pm )
+ J_{4 \mu}( \pm, \pm, \mp, \pm ) + J_{6 \mu}(\pm, \pm, \mp, \pm )
\;\; ,
$$

$$
\begin {array}{c} \displaystyle
                        \\[0.5cm] \displaystyle
J_{2 \mu}( \pm, \pm, \mp, \pm )
= { 4 N_1 N_2 (q q') \over k_1 k_2 - q k_1 - q k_2 }
{ {\it i} \over [ q_0 q'_0 (q p) (q' p) ]^{1/2} }
                        \\[0.5cm] \displaystyle
\Bigl\{ p_{\mu} [ ( q' k_2 - q q' )
\{ (q k_1) (q' k_2) - (q q') (k_1 k_2) \}
- (q k_2) (q' k_1) (q' k_2) - (q q') (q k_2) (q' k_1) ]
                        \\[0.5cm] \displaystyle
+ q_{\mu} [ ( q' k_2 - q q' ) \{ (q' k_1) (p k_2)
- (q' k_2) (p k_1) + (q' p) (k_1 k_2) \}
+ 2 (q k_2) (q' k_1) (q' p) ]
                        \\[0.5cm] \displaystyle
+ {q'}_{\mu} [ ( q q' + q' k_2 )
\{ (q k_2) (p k_1) - (q k_1)(p k_2) - (q p) (k_1 k_2) \}
\mp 2 (q k_2) \varepsilon (q, q', p, k_1)
                        \\[0.5cm] \displaystyle
+ 2 (q k_1) \{ (q' k_2)(qp)
\pm \varepsilon (q, q', p, k_2) \}
+ 2 (k_1 k_2) \{ (q q') (p k_2) - (q' p) (q k_2)
\mp \varepsilon (q, q', p, k_2) \} ]
                        \\[0.5cm] \displaystyle
+ k_{1 \mu}( q' k_2 - q q' )
[ (q p) (q' k_2 ) - (q q') (p k_2) + (q k_2) (q' p)
\pm \varepsilon (q, q', p, k_2 ) ]
                        \\[0.5cm] \displaystyle
- k_{2 \mu} ( q' k_2 - q q' )
[ (q p) (q' k_1) - (q q') (p k_1) + (q k_1) (q' p)
\pm \varepsilon (q, q', p, k_1 ) ]
                        \\[0.5cm] \displaystyle
\mp ( q' k_2 - q q' ) [ (k_1 k_2) \varepsilon (\mu, q, q', p)
+ (q q') \varepsilon (\mu, p, k_1, k_2)
                        \\[0.5cm] \displaystyle
- (q p)  \varepsilon (\mu,q', k_1, k_2)
- (q' p) \varepsilon (\mu, q, k_1, k_2) ]
\mp 2 (q k_2) (q' k_1) \varepsilon (\mu, q, q', p) \Bigr\}
\;\; ,
\end {array}
$$

$$
\begin {array}{c} \displaystyle
J_{4 \mu}(\pm, \pm, \mp, \pm)
= 4 N_1 N_2 (q q')
{ {\it i} \over [ q_0 q'_0 (q p) (q' p) ]^{1/2} }
                        \\[0.5cm] \displaystyle
\Bigl\{ p_{\mu} [ 2 (q q') ( -q q' - q k_1 + q' k_2 )
+ (q q') (k_1 k_2) + (q k_1) (q' k_2) - (q k_2) (q' k_1) ]
                        \\[0.5cm] \displaystyle
+ q_{\mu} [ 2 (q q') ( q' p + p k_1 ) - 2 (q p) (q' k_1)
- 2 (q' p)(q' k_2)
+ (q' k_1) (p k_2) - (p k_1) (q' k_2) - (q' p) (k_1 k_2) ]
                        \\[0.5cm] \displaystyle
+ {q'}_{\mu} [ 2 (q q') ( q p - p k_2 ) + 2 (q p) (q k_1)
+ 2 (q' p) (q k_2)
- (q k_1) (p k_2) + (p k_1) (q k_2) - (q p) (k_1 k_2) ]
                        \\[0.5cm] \displaystyle
+ k_{1 \mu} [ (q p) (q' k_2) + (q' p) (q k_2)
- (q q') (p k_2) \pm \varepsilon (q, q', p, k_2) ]
                        \\[0.5cm] \displaystyle
+ k_{2 \mu} [ (q p) (q' k_1) + (q' p) (q k_1)
- (q q') (p k_1) \pm \varepsilon (q, q', p, k_1) ]
                        \\[0.5cm] \displaystyle
\mp 2 (q q') [ \varepsilon (\mu, q, q', p)
- \varepsilon (\mu, q, p, k_1) - \varepsilon (\mu, q', p, k_2) ]
                        \\[0.5cm] \displaystyle
\mp 2 (q p) \varepsilon (\mu, q, q', k_1)
\pm 2 (q' p) \varepsilon (\mu, q, q', k_2)
\pm (k_1 k_2) \varepsilon (\mu, q, q',p)
                        \\[0.5cm] \displaystyle
\mp (q q') \varepsilon (\mu, p, k_1, k_2)
\pm (q p) \varepsilon (\mu, q', k_1, k_2)
\pm (q' p) \varepsilon (\mu, q, k_1, k_2) \Bigr\}
\;\; ,
\end {array}
$$

$$
\begin {array}{c} \displaystyle
J_{6 \mu}(\pm, \pm, \mp, \pm)
= { 4 N_1 N_2 (q q') \over k_1 k_2 + q' k_1 + q' k_2 }
 { {\it i} \over [ q_0 q'_0 (q p) (q' p) ]^{1/2} }
                        \\[0.5cm] \displaystyle
\Bigl\{ p_{\mu} [ ( q q' + q k_1 ) \{ (q q') (k_1 k_2)
- (q k_1) (q' k_2) \}
+ (q k_1) (q k_2) (q' k_1)
- (q q') (q k_2) (q' k_1) ]
                        \\[0.5cm] \displaystyle
+ q_{\mu} [ ( q q' - q k_1 ) \{ (q' k_1) (p k_2)
- (q' k_2) (p k_1) - (q' p) (k_1 k_2) \}
\mp 2 (q' k_1) \varepsilon (q, q', p, k_2)
                        \\[0.5cm] \displaystyle
+ 2 (q' k_2) \{ (q k_1) (q' p)
\pm \varepsilon (q, q', p, k_1) \}
+ 2(k_1 k_2) \{ (q p) (q' k_1) - (q q') (p k_1)
\pm \varepsilon (q, q', p, k_1) \} ]
                        \\[0.5cm] \displaystyle
+ {q'}_{\mu} [ ( q q'+ q k_1 ) \{ (q k_1) (p k_2)
- (q k_2) (p k_1) - (q p) (k_1 k_2) \}
+ 2 (q k_2) (q' k_1) (q p) ]
                        \\[0.5cm] \displaystyle
+ k_{1 \mu} ( q q' + q k_1 ) [ (q' p) (q k_2) - (q q') (p k_2)
+ (q p) (q' k_2) \pm \varepsilon (q, q', p, k_2) ]
                        \\[0.5cm] \displaystyle
- k_{2 \mu} ( q q' + q k_1 ) [ (q'p) (q k_1) - (q q') (p k_1)
+ (q p) (q' k_1) \pm \varepsilon (q, q', p, k_1) ]
                        \\[0.5cm] \displaystyle
\pm ( q q' + q k_1) [ (k_1 k_2) \varepsilon (\mu, q, q', p)
+ (q q') \varepsilon (\mu, p, k_1, k_2)
                        \\[0.5cm] \displaystyle
- (q p) \varepsilon (\mu, q', k_1, k_2)
- (q' p) \varepsilon (\mu, q, k_1, k_2) ]
\mp  2 (q k_2) (q' k_1) \varepsilon (\mu, q, q', p) \Bigr\}
\;\; .
\end {array}
$$

   The expression for
$$
\displaystyle
J_{\mu}(\pm ,\mp ,\pm ,\pm) = J_{1 \mu}(\pm ,\mp ,\pm ,\pm )
+ J_{3 \mu}(\pm ,\mp ,\pm ,\pm) + J_{5 \mu}(\pm ,\mp ,\pm ,\pm)
$$
is obtained from the one for
$
\displaystyle
J_{\mu}(\pm ,\pm ,\mp ,\pm)
\,
$
by interchanging $k_1$ and $k_2$.
In all expressions, $p$ is an arbitrary $4$-momentum such that
$
\displaystyle
p^2 = 0
\; .
$

    Our expressions for leptonic currents are fairly compact.
For comparison, we note that the leptonic tensors

$
\displaystyle
\,
J_{\mu}(\pm, \mp, \pm, \pm) {J_{\nu}}^{*}(\pm, \mp, \pm, \pm)
\,
$
and
$
\displaystyle
\,
J_{\mu}(\pm, \pm, \mp,\pm) {J_{\nu}}^{*}(\pm, \pm, \mp, \pm)
\,
$  \\
each contain $1416$ terms when calculated by the classical method
by means of the computer system SCHOONSCHIP.

   I thank S.M.Sikach and Professor N.M.Shumeiko for numerous
helpful discussions during work on this paper.

%%%%%%%%%%%%%%%%%%%%%%%%%%%%%%%%%%%%%%%%%%%%%%%%%%%%%%%%%%%%%%%%%%%%%%%%%
%%%%%%%%%%%%%%%%%%%%%%%%%%%%%%%%%%%%%%%%%%%%%%%%%%%%%%%%%%%%%%%%%%%%%%%%%

%REFERENCES

\end {document}